\documentstyle [preprint,aps]{revtex}
\begin{document}
\draft
\title{The Korteweg-de Vries Hierarchy and Long Water-Waves}
\author{R. A. Kraenkel\cite{kra} and M. A. Manna }
\address{Physique Math\'ematique et Th\'eorique, URA-CNRS 768\\
Universit\'e de Montpellier II\\
34095 Montpellier Cedex 05\\
France}
\author{J. G. Pereira }
\address{Instituto de F\'{\i}sica Te\'orica\\
Universidade Estadual Paulista\\
Rua Pamplona 145\\
01405-900\, S\~ao Paulo\, SP --
Brazil}
\maketitle
\begin{abstract}
By using the multiple scale method with the simultaneous introduction
of multiple times, we study the propagation of long surface-waves in a
shallow inviscid fluid. As a consequence of the
requirements of scale invariance and absence of secular terms in each order of
the perturbative expansion, we show that the Korteweg-de Vries hierarchy
equations do appear in the description of such waves.
Finally, we show that this procedure of eliminating secularities is
closely related to the renormalization technique introduced by Kodama and
Taniuti.
\end{abstract}
\pacs{03.40.Kf \, ; \, 47.35.+i \, ; \, 03.40.Gc}

\vfill \eject
\section{Introduction}

In $1967$, Gardner, Greene, Kruskal and Miura \cite{ggkm}, by making use of the
ideas of direct and inverse scattering, showed that the Korteweg-de Vries (KdV)
equation could be solved exactly as an initial value problem. Shortly after,
Lax \cite{lax} generalized these ideas to a general evolution equation of the
type
\begin{equation}
\frac{\partial u}{\partial t} = K[u] \, ,
\label{eq 1}
\end{equation}
with $K$ a nonlinear operator that could be written in the form
\begin{equation}
K[u] = \left[ B,L \right] \, ,
\label{eq 2}
\end{equation}
and with $B$ and $L$ self--adjoint linear operators depending on $u$. As
a consequence of this formalism, Lax showed the existence of an infinite
sequence of integrable partial differential equations of the form
\begin{equation}
\frac{\partial u}{\partial t} = K_n[u] \quad , \quad n = 1, 2, ... \, ,
\label{eq 3}
\end{equation}
where
\begin{equation}
K_n =\left[ B_n , L \right]
\label{eq 4}
\end{equation}
with
\begin{equation}
B_n = D^{2n+1} + \sum_{j=1}^n \left(b_j D^{2j+1} + D^{2j+1} b_j \right)
\label{eq 5}
\end{equation}
and
\begin{equation}
L = D^2 + \frac{1}{6} u \, .
\label{eq 6}
\end{equation}
In these equations, $b_j$ are coefficients that can be determined
\cite{lax}, and $D = {\partial} / {\partial x}$.
For $n=1$, the KdV equation is obtained. For $n \geq 2$, we have the higher
order KdV equations.
This infinite sequence of integrable nonlinear partial differential
equations form the so called KdV hierarchy \cite{lax}. Along with
KdV, they are all integrable by the inverse scattering method, and they have
the
same integrals of motion as KdV. However, in contrast to KdV, which has been
shown to govern the nonlinear long-wave dynamics of general dispersive systems,
the physical relevance of the higher order
equations of the KdV hierarchy has remained, up to now, obscure.

We will show in this paper that the equations of the KdV hierarchy do appear
in the description of
physical systems. More specifically, we will first argue that the same
amplitude that satisfies the KdV equation,
also satisfies  the higher order equations of the KdV hierarchy, each one in
a different time-scale. As a consequence, we will then show that this property
prevents divergent solutions to the physical problem. The infinite sequence of
time-scales are defined from the basic time variable $t$ by $\tau_1 = \epsilon
^{\frac{1}{2}} t$, $\tau_3 =
\epsilon^{\frac{2}{3}} t$, $\tau_5 = \epsilon^{\frac{5}{2}} t$,..., where
$\epsilon$ is a parameter satisfying
$\epsilon \ll 1$. The primary reason for
introducing them  is that they may be used as a tool to eliminate the
secularities that appear in the
problem when a perturbative solution for the higher order terms of the
amplitude is searched. However, when a function of a single time variable
$F(t)$
is extended to a function of several time variables $F(\tau_1 ,\tau_3 ,\tau_5
,..)$, the problem of knowing
the allowed evolution in each one of the different time-scales is posed
\cite{sandri}. Indeed, as we are going to see, this evolution
is not arbritary. It is determined by a scale invariance requirement, which
must hold to ensure the
ordering of the expansion, and by the compatibility condition
$$
\frac{\partial^2 F}{\partial \tau_{3}
\partial \tau_{2n+1}} = \frac{\partial^2 F}{\partial \tau_{2n+1} \partial
\tau_{3}} \, .
$$
If, for a physical system, the evolution of a certain amplitude is governed by
the KdV equation in the time $\tau_3$, we will
show that the above compatibility condition constraints the evolution on higher
order times of that amplitude to be
governed by the higher order equations of the KdV hierarchy, leaving only one
free parameter at each order which is related to
the possibility of redifining the time $\tau_{2n+1}$ by a multiplicative factor
$\alpha_{2n+1}$. By choosing specific values for the free parameters, the task
of secularity elimination can then be accomplished. As a consequence, our
construction of a {\it bona fide} perturbative expansion will provide a link
between the equations of the
KdV hierarchy and the evolution of a physical quantity. The physical system
studied here, that of long surface water-- waves, is a kind of classical system
where our ideas are well
illustrated. However, the results obtained are, to a certain extend, model
independent, and in this sense the study made here can be considered as
representative of a general physical situation.

The paper is organized as follows. In section II, we
obtain the basic equations describing surface water waves
in the multiple space and time formalism. In section III, the first three
evolution equations are obtained.
In section IV, by using the multiple time formalism,
we examine how the symmetry of the time derivatives determines the evolution of
the wave amplitude in any time
scale.
The use of the KdV hierarchy equations to eliminate the secularities in the
evolution of the
higher order terms of the wave amplitude is discussed in section V. In section
VI, we consider the case of
a traveling-wave
solution to the equations of the KdV hierarchy, and we show that the method of
eliminating secularities by using these equations can be related to the results
of Kodama
and Taniuti~\cite{kota}, where a renormalization technique was
introduced to obtain a secular free perturbative
expansion. And finally, in section VII, we summarize and discuss the results
obtained.

\section{The Multiple Space and Time Formalism for Water Waves}

We consider a two-dimensional inviscid incompressible fluid in a constant
gravitational field. The space coordinates are denoted by $(x,z)$ and the
corresponding components of the velocity $\vec v$ by $(u,w)$. The gravitational
acceleration $\vec g$ is in the negative $z$ direction. The equations
describing such a fluid are \cite{whit}: the incompressibility equation
\begin{equation}
\vec{\nabla} \cdot \vec{v} = 0 \, ,
\label{eq 32}
\end{equation}
and the Euler equation
\begin{equation}
\frac{D \vec v}{D t} \equiv \frac{\partial \vec v}{\partial t} + \left(\vec v
\cdot \vec{\nabla} \right) \vec v = - \frac{1}{\rho} \vec{\nabla} p +
\vec g \, ,
\label{eq 33}
\end{equation}
where $\rho$ is the fluid density, $p$ is the presure and $\hat k$ is an unit
vector in the $z$ direction. If we assume the fluid to be irrotational,
\begin{equation}
\vec{\nabla} \times \vec v = 0 \, ,
\label{eq 34}
\end{equation}
and consequently a velocity potential $\phi$ can be introduced through
\begin{equation}
\vec v = \vec{\nabla} \phi \, .
\label{eq 35}
\end{equation}
Using this definition, Eq.(\ref{eq 32}) becomes
\begin{equation}
{\nabla}^2 \phi = 0 \, ,
\label{eq 36}
\end{equation}
while Eq.(\ref{eq 33}), after an integration, reads
\begin{equation}
\frac{p - p_0}{\rho} = - \frac{\partial}{\partial t} \phi - \frac{1}{2}
\left(\vec{\nabla} \phi \right)^2 - g z \, ,
\label{eq 37}
\end{equation}
with $p_0$ an integration constant.

We consider the case of a fluid of height $h$, limited above by a passive gas
exerting a constant pressure $p_0$ on it, and let the upper surface to be
described by
\begin{equation}
z = \zeta (x,t) \, .
\label{eq 38}
\end{equation}
The kinematic boundary condition at this surface is then written in the form
\begin{equation}
\frac{D \zeta}{D t} \equiv \zeta_t + \phi_x \zeta_x = \phi_z \, ,
\label{eq 39}
\end{equation}
with the subscripts denoting partial derivatives.
However, there is also a dynamic boundary condition obtained from Eq.(\ref{eq
37}), which reads
\begin{equation}
\phi_t + \frac{1}{2} \left[{(\phi_x)}^2 + {(\phi_z)}^2 \right] + g \rho = 0 \,
,
\label{eq 40}
\end{equation}
on $z = \zeta(x,t)$. Finally, the lower boundary is supposed to be a rigid
horizontal flat botton, localized at $z = - h$. In this case, the corresponding
boundary condition implies that the normal velocity of the fluid must vanish:
\begin{equation}
\phi_z = 0 \, .
\label{eq 41}
\end{equation}

We now wish to consider the long-wave in shallow-water  approximation to the
above equations. This may be
done, for instance, by using the the reductive perturbation method of Taniuti
\cite{taniu}, which
introduces slow space and time variables. The slow space variable is given by:
\begin{equation}
\xi^{\prime} = \epsilon^{\frac{1}{2}} x \, .
\label{eq 42}
\end{equation}
Next, we introduce a very general set of slow time variables \cite{jeka}:
\begin{equation}
\tau_1 = \epsilon^{\frac{1}{2}} t \; , \; \tau_3 = \epsilon^{\frac{3}{2}} t \;
, \; \tau_5 = \epsilon^{\frac{5}{2}} t \; , \ldots \; .
\label{eq 43}
\end{equation}
In addition, we expand $\zeta$ and $\phi$ in a suitable power series in the
parameter $\epsilon$:
\begin{equation}
\zeta = \epsilon \hat{\zeta} \equiv \epsilon \left(\zeta_0 + \epsilon \zeta_1 +
\epsilon^2 \zeta_2 + \ldots \right)
\label{eq 44}
\end{equation}
\begin{equation}
\phi = \epsilon^{\frac{1}{2}} \hat{\phi} \equiv \epsilon^{\frac{1}{2}}
\left(\phi_0 + \epsilon \phi_1 +
\epsilon^2 \phi_2 + \ldots \right) \, .
\label{eq 45}
\end{equation}
Introducing these expansions and the slow variables into the water wave
equations, we obtain:
\begin{equation}
\epsilon {\hat \phi}_{\xi^\prime \xi^\prime} + {\hat \phi}_{zz} = 0 \quad ,
\quad -h < z < \epsilon \hat \zeta
\label{eq 46}
\end{equation}
\begin{equation}
{\hat \phi}_z = 0 \quad , \quad z = - h
\label{eq 47}
\end{equation}
\begin{equation}
{\hat \phi}_z = \epsilon {\hat \zeta}_{\tau_1} + \epsilon^2 {\hat
\zeta}_{\tau_3} + \epsilon^3 {\hat \zeta}_{\tau_5} + \ldots + \epsilon^2 {\hat
\phi}_{\xi^\prime} {\hat \zeta}_{\xi^\prime} \quad , \quad z = \epsilon \hat
\zeta
\label{eq 48}
\end{equation}
\begin{equation}
2 g \hat \zeta + 2 {\hat \phi}_{\tau_1} + 2 \epsilon {\hat \phi}_{\tau_3} + 2
\epsilon^2 {\hat \phi}_{\tau_5} + \ldots + \epsilon {\hat \phi}_{\xi^\prime}
{\hat \phi}_{\xi^\prime} + {\hat \phi}_z {\hat \phi}_z = 0 \quad , \quad z =
\epsilon \hat \zeta \, .
\label{eq 49}
\end{equation}
These are the basic equations we are going to use to obtain the evolution
equations describing the system.

\section{The First Evolution Equations}

Equation (\ref{eq 46}) for the velocity potential can be solved for the
boundary condition (\ref{eq 47}), independently of equations
(\ref{eq 48}) e (\ref{eq 49}). At order
$\epsilon^0$, the solution is
\begin{equation}
\phi_0 = {\cal F} \, ,
\label{eq 50}
\end{equation}
with ${\cal F} = {\cal F}(\xi^\prime;\tau_1,\tau_3,\tau_5,\ldots)$ an arbitrary
function. At order $\epsilon^1$, it is given by
\begin{equation}
\phi_1 = - \left(\frac{z^2}{2} + h z \right) {\cal F}_{\xi^\prime \xi^\prime} +
{\cal G} \, ,
\label{eq 51}
\end{equation}
with ${\cal G} = {\cal G}(\xi^\prime;\tau_1,\tau_3,\tau_5,\ldots)$ another
arbitrary function. At order $\epsilon^2$, we get
\begin{equation}
\phi_2 = - \frac{1}{24} \left(z^4 + 4 h z^3 - 8 h^3 z \right) {\cal F}_{(4
\xi^\prime)} - \frac{1}{2} \left(z^2 + 2 h z \right){\cal G}_{\xi^\prime
\xi^\prime} +
{\cal H} \, ,
\label{eq 52}
\end{equation}
with ${\cal H} = {\cal H}(\xi^\prime;\tau_1,\tau_3,\tau_5,\ldots)$ another
arbitrary function. At order $\epsilon^3$, the solution
is
\begin{eqnarray}
\phi_3 = &-& \frac{1}{720} \left(z^6 + 6 h z^5 - 40 h^3 z^3 + 16 h^6 \right)
{\cal F}_{(6 \xi^\prime)} \nonumber \\
&+& \frac{1}{24} \left(z^4 + 4 h z^3 - 8 h^3 z \right) {\cal G}_{(4\xi^\prime)}
- \frac{1}{2} \left(z^2 + 2 h z \right) {\cal H}_{\xi^\prime \xi^\prime} +
{\cal I} \, ,
\label{eq 53}
\end{eqnarray}
with ${\cal I} = {\cal I}(\xi^\prime;\tau_1,\tau_3,\tau_5,\ldots)$ again an
arbitrary function. We could proceed further and
calculate $\phi_4, \phi_5, \ldots \; $. But, we stop here since this is all we
are going to need.

We pass now to eqs.(\ref{eq 48}) and (\ref{eq 49}), and solve the first few
orders in our perturbative scheme. At order $\epsilon^0$,
Eq.(\ref{eq 48}) gives nothing, while Eq.(\ref{eq 49}) reads
\begin{equation}
g \zeta_0 + \phi_{0\tau_1} = 0 \, .
\label{eq 54}
\end{equation}
Substituting $\phi_0$, we get
\begin{equation}
\zeta_0 = - \frac{1}{g} {\cal F}_{\tau_1} \, .
\label{eq 55}
\end{equation}
At order $\epsilon^1$, Eq.(\ref{eq 48}) is
\begin{equation}
{\cal F}_{\tau_1\tau_1} - g h {\cal F}_{\xi^\prime \xi^\prime} = 0 \, .
\label{eq 56}
\end{equation}
Defining
\begin{equation}
g h \equiv c^2 \, ,
\label{eq 57}
\end{equation}
with $c$ a velocity, its solution can be written in the form
$$
f = F(\xi^\prime - c \tau_1) + G(\xi^\prime + c \tau_1) \, ,
$$
where the sign $- (+)$ refers to a wave moving to the right (left) with a
velocity $c$. For definiteness, let us choose the wave moving to the right, and
let us define a new
coordinate system by
\begin{equation}
\xi = \xi^\prime - c \tau_1 \, .
\label{eq 58}
\end{equation}
Therefore,
\begin{equation}
{\cal F}_{\tau_1} = - c {\cal F}_{\xi} \, ,
\label{eq 59}
\end{equation}
and from Eq.(\ref{eq 55}) we get
\begin{equation}
\zeta_0 = \frac{h}{c} {\cal F}_{\xi} \, .
\label{eq 60}
\end{equation}
Consequently,
\begin{equation}
\zeta_{0\tau_1} + c \zeta_{0\xi} = 0 \, ,
\label{eq 61}
\end{equation}
which is the usual linear wave equation.

Now, using eqs.(\ref{eq 42}) and (\ref{eq 43}), we can rewrite Eq.(\ref{eq 58})
in the form:
\begin{equation}
\xi = \epsilon^{\frac{1}{2}} \left(x - c t\right) \, ,
\label{eq 62}
\end{equation}
from which we see that $\xi$ is a coordinate system moving with a velocity $c$
of the linear waves, in relation to the laboratory coordinate $x$.
Therefore, from now on, all dependent variables will be assumed to depend on
$(\xi^\prime,\tau_1)$ through $\xi$ only, which automatically implies that they
satisfiy Eq.(\ref{eq 59}).

At order $\epsilon^1$, Eq.(\ref{eq 49}) gives
\begin{equation}
2 g \zeta_1 - 2 c {\cal G}_{\xi} + 2 {\cal F}_{\tau_3} + {\cal F}_{\xi} {\cal
F}_{\xi} = 0 \, ,
\label{eq 63}
\end{equation}
where we have already used
Eq.(\ref{eq 59}) for $\cal G$. Derivating in relation to $\xi$, and using
Eq.(\ref{eq 60}), we obtain
\begin{equation}
c \zeta_{1\xi} - h {\cal G}_{\xi\xi} = - \zeta_{0\tau_3} - \frac{c}{h} \zeta_0
\zeta_{0\xi} \, .
\label{eq 64}
\end{equation}

Now we go to the order $\epsilon^2$. From Eq.(\ref{eq 48}), we have
\begin{equation}
c \zeta_{1\xi} - h {\cal G}_{\xi\xi} = \zeta_{0\tau_3} + 2 \frac{c}{h} \zeta_0
\zeta_{0\xi} + \frac{c h^2}{3} \zeta_{0\xi\xi\xi} \, .
\label{eq 65}
\end{equation}
Substituting into Eq.(\ref{eq 64}), we get
\begin{equation}
\zeta_{0\tau_3} + \frac{3 c}{2 h} \zeta_0 \zeta_{0\xi} + \frac{c h^2}{6}
\zeta_{0\xi\xi\xi} = 0 \, ,
\label{eq 66}
\end{equation}
which is the KdV equation. It is the first nontrivial equation in the KdV
hierarchy. At this point we can see the importance of introducing multiple
times: while the linear wave equation (\ref{eq 61}) describes the evolution
of $\zeta_0$ in the time $\tau_1$, the KdV equation describes the evolution of
$\zeta_0$ in the time $\tau_3$. In other words, the phenomena described by each
one of these equations occur in different time scales. And this will be the
case for all the higher order equations of the KdV hierarchy.
Now, from Eq.(\ref{eq 49}) at order $\epsilon^2$, we obtain
\begin{equation}
2 g \zeta_2 + 2 c^2 \zeta_0 \zeta_{0\xi\xi} - 2 c {\cal H}_{\xi} + 2 {\cal
G}_{\tau_3} +
2 {\cal F}_{\tau_5} + 2 {\cal F}_{\xi} {\cal G}_{\xi} + c^2 \zeta_{0\xi}
\zeta_{0\xi} = 0 \, .
\label{eq 67}
\end{equation}
Derivating in relation to $\xi$, and using Eq.(\ref{eq 60}), it becomes
\begin{equation}
c \zeta_{2\xi} - h {\cal H}_{\xi\xi} + \frac{h}{c} {\cal G}_{\xi\tau_3} +
\left(
\zeta_{0\xi} {\cal G}_{\xi} + \zeta_0 {\cal G}_{\xi\xi} \right) = -
\zeta_{0\tau_5} - 2 c h
\zeta_{0\xi} \zeta_{0 \xi\xi} - c h \zeta_0 \zeta_{0\xi\xi\xi} \, .
\label{eq 68}
\end{equation}

We pass now to the order $\epsilon^3$. From Eq.(\ref{eq 48}), we have
\begin{eqnarray}
c \zeta_{2\xi} - h {\cal H}_{\xi} - \left(\zeta_{0} {\cal G}_{\xi\xi} +
\zeta_{0\xi} {\cal G}_{\xi} \right) &{}& \nonumber \\
- \frac{c}{h} \left(\zeta_{0\xi} \zeta_1 + \zeta_0 \zeta_{1\xi}\right)
&-& \frac{h^3}{3} {\cal G}_{(4\xi)}
= \zeta_{0\tau_5} + \frac{2}{15} c h^4 \zeta_{0(5\xi)} \, .
\label{eq 69}
\end{eqnarray}
Equations (\ref{eq 68}) and (\ref{eq 69}) can be combined to yield an evolution
equation involving $\zeta_{1\tau_3}$, $\zeta_{0\tau_5}$ and ${\cal G}_{\xi
\tau_3}$:
\begin{eqnarray}
2 \zeta_{0\tau_5} &+& \frac{2}{15} c h^4 \zeta_{0(5\xi)} + 2 c h \zeta_{0\xi}
\zeta_{0\xi\xi} + c h \zeta_0 \zeta_{0\xi\xi\xi} + \zeta_{1\tau_3} \nonumber \\
&+& \frac{c}{h} \left(\zeta_{0\xi} \zeta_1 + \zeta_0 \zeta_{1\xi} \right)
+ \frac{h}{c} {\cal G}_{\xi \tau_3} + 2 \left( \zeta_0 {\cal G}_{\xi\xi} +
\zeta_{0\xi} {\cal G}_{\xi} \right) + \frac{h^3}{3} {\cal G}_{(4\xi)} = 0 \, .
\label{eq 70}
\end{eqnarray}
Now, making use of Eq.(\ref{eq 66}) to describe ${\cal F}_{\tau_3}$,
Eq.(\ref{eq 63}) can be put in the form
\begin{equation}
\zeta_1 - \frac{h}{c} {\cal G}_{\xi} = \frac{1}{4 h} {\zeta_0}^2 +
\frac{h^2}{6}
\zeta_{0\xi\xi} \, .
\label{eq 71}
\end{equation}
We can then use this equation to eliminate ${\cal G}_{\xi}$ from Eq.(\ref{eq
70}). The result is
\begin{equation}
\zeta_{1\tau_3} + \frac{3 c}{2 h} \left(\zeta_0 \zeta_1 \right)_{\xi} + \frac{c
h^2}{6} \zeta_{1\xi\xi\xi} = S(\zeta_0) \, ,
\label{eq 72}
\end{equation}
where
\begin{equation}
S(\zeta_0) = - \zeta_{0\tau_5} - \frac{19}{360} c h^4 \zeta_{0(5\xi)} -
\frac{5}{12} c h \zeta_0 \zeta_{0\xi\xi\xi} - \frac{23}{24} c h \zeta_{0\xi}
\zeta_{0\xi\xi} + \frac{3 c}{8h^2} {\zeta_0}^2 \zeta_{0\xi} \, .
\label{eq 73}
\end{equation}
Equation (\ref{eq 72}), as it stands, can not be viewed as an evolution
equation for $\zeta_1$ in the time $\tau_3$ because the non--homogeneous term
involves
$\zeta_{0\tau_5}$, and the evolution of $\zeta_{0}$ in $\tau_5$
is not known up to this point. Moreover, the term proportional to
$\zeta_{0(5\xi)}$ in
$S(\zeta_0)$ is a resonant term, that is, it is a secular producing term to
the solution $\zeta_1$ \cite{kota}. We will show in the next sections how the
evolution of $\zeta_{0}$ in $\tau_5$ may be used to cancel the secular term.
However, let us first state a mathematical result related to the KdV hierarchy,
which will be used later.

\section{The Symmetry of Time Derivatives and the \protect\\
Korteweg--de Vries Hierarchy}

As is widely known \cite{gard}, the KdV equation
\begin{equation}
\zeta_t = 6 \zeta \zeta_x - \zeta_{xxx} \, ,
\label{eq 7}
\end{equation}
is a Hamiltonian system, and can be written in the form
\begin{equation}
\zeta_t = \frac{\partial}{\partial x} \left(\frac{\delta H}{\delta \zeta}
\right) \, ,
\label{eq 8}
\end{equation}
with the Hamiltonian $H$ given by
\begin{equation}
H = \int\limits_{- \infty}^{+ \infty} \left[ \frac{(\zeta_x)^2}{2} + \zeta^3
\right] dx
\, .
\label{eq 9}
\end{equation}

On the other hand, as is also well known, the KdV equation has an infinite
sequence
of conservation laws \cite{miura,mgk}, with the integrals of motion given by
\begin{equation}
I_n[\zeta] = \int\limits_{- \infty}^{+ \infty} T_n dx \quad , \quad
n = 0,1,2,3,\ldots \; \; ,
\label{eq 10}
\end{equation}
where $T_n$, the conserved density, is a polynomial in $\zeta, \zeta_x,
\zeta_{xx}$, etc. Their explicit form can be obtained through a perturbative
scheme developed by Miura, Gardner and Kruskal \cite{mgk}, which is based on
expansions on a small parameter $\epsilon$. In this method, each integral of
motion $I_n$ appears in a different order of the perturbation parameter
$\epsilon$. Now, the infinite sequence of equations of
the KdV hierarchy can be expressed in terms of these integrals of
motion according to \cite{novik}
\begin{equation}
\zeta_t = \frac{\partial}{\partial x} \left( \frac{\delta I_n}{\delta \zeta}
\right) \, .
\label{eq 11}
\end{equation}
For $n = 1$, the integral $I_1$ coincides with the Hamiltonian given by
Eq.(\ref{eq 9}), and the KdV
equation is obtained, which is the first nontrivial member of the KdV
hierarchy. From the explicit form of $I_n$ \cite{mgk}, it is easy to show
that each one of Eqs.(\ref{eq 11}) is invariant, up to a Galilean term,
under the transformations
\begin{equation}
\xi = \epsilon^{\frac{1}{2}} (x - c t)
\label{eq 12}
\end{equation}
\begin{equation}
\tau_{2n+1} = \epsilon^{n + \frac{1}{2}} \; t \quad ; \quad n = 1,2,3,\ldots
\label{eq 13}
\end{equation}
\begin{equation}
\zeta = \epsilon \, \zeta_0 \, .
\label{eq 14}
\end{equation}
In these new coordinates, called slow variables, Eq.(\ref{eq 11}) is written
in the form
\begin{equation}
\zeta_{0\tau_{2n+1}} = \frac{\partial}{\partial \xi} \left( \frac{\delta
I_n}{\delta \zeta_0} \right) \; .
\label{eq 15}
\end{equation}
Notice that an infinite sequence of slow time variables $\tau_3$, $\tau_5$,
$\tau_7$, $\ldots \,$ was introduced in the above transformations. As a
consequence, any dependent variable $F(x;t)$ will be
a function of these multiple times:
\begin{equation}
F(x;t) = F(x;\tau_1,\tau_3,\tau_5,\ldots) \, .
\label{eq 16}
\end{equation}
Therefore, the time derivative of $F(x;t)$ turns out to be
\begin{equation}
\frac{\partial F}{\partial t} = \left(\epsilon^{\frac{3}{2}}
\frac{\partial}{\partial \tau_3} + \epsilon^{\frac{5}{2}}
\frac{\partial}{\partial \tau_5} + \epsilon^{\frac{7}{2}}
\frac{\partial}{\partial \tau_7} + \ldots \right) F \, .
\label{eq 17}
\end{equation}

Let us now return to our physical system. As we have already seen, the
evolution of $\zeta_0$ in the time $\tau_3$ was found to be the KdV equation
\begin{equation}
\zeta_{0\tau_3} = \alpha_1 \zeta_{0\xi\xi\xi} + \beta_1 \zeta_0 \zeta_{0\xi}
\, ,
\label{eq 18}
\end{equation}
with
\begin{equation}
\alpha_1 = \frac{c h^2}{6} \qquad , \qquad \beta_1 = \frac{3 c }{2 h} \, .
\label{eq }
\end{equation}
Let us now examine how the evolution of $\zeta_0$ in times $\tau_5 ,
\tau_7$,... can be obtained.
 The first crucial point of this analysis is to note that, for the evolution of
$\zeta_0$ in the time $\tau_3$, the two terms in the r.h.s. of Eq.(\ref{eq 18})
exhaust
all possible terms such that, when transforming back from
the slow to the laboratory variables,
the resulting equation does not depend on $\epsilon$.

In the very same way, the evolution equation of
$\zeta_0$ in the time $\tau_5$ must only involve terms presenting the same
property as the terms in
the KdV equation. If this where not so, the ordering of the perturbative series
would not be ensured.
In other words, this evolution equation must be formally the
same in the slow $(\xi,\tau_5,\zeta_0)$ as well as in the laboratory
coordinates
$(x,t,\zeta)$, establishing thus a scale invariance requirement. A simple
analysis shows that the only possible terms are:
\begin{equation}
\zeta_{0\tau_5} = \alpha_2 \zeta_{0(5\xi)} + \beta_2 \zeta_0 \zeta_{0\xi\xi\xi}
+ \left(\beta_2 + \gamma_2 \right) \zeta_{0\xi} \zeta_{0\xi\xi} + \delta_2
{\zeta_0}^2 \zeta_{0\xi} \, ,
\label{eq 19}
\end{equation}
where $\alpha_2, \beta_2, \gamma_2, \delta_2$ are constants. Now comes the
second crucial point: the coefficients $\alpha_2, \beta_2, \gamma_2, \delta_2$
are not completely arbitrary if $\zeta_0$ is to satisfy also the KdV equation
in the time $\tau_3$.
Instead, they are constrainted by the relations arising when the natural (in
the multiple time formalism) compatibility condition
\begin{equation}
\left(\zeta_{0\tau_3}\right)_{\tau_5} = \left(\zeta_{0\tau_5}\right)_{\tau_3}
\label{eq 20}
\end{equation}
is imposed. This condition only makes sense in the multiple time formalism,
since otherwise it would be redundant and would lead to a trivial identity. A
tedious but straightforward calculation shows that the relations arising from
this condition are:
\begin{equation}
\frac{\beta_2}{\alpha_2} = \frac{5}{3} \frac{\beta_1}{\alpha_1} \quad , \quad
\frac{\gamma_2}{\alpha_2} = \frac{5}{3} \frac{\beta_1}{\alpha_1} \quad , \quad
\frac{\delta_2}{\alpha_2} = \frac{5}{6} \left(\frac{\beta_1}{\alpha_1}
\right)^2 \, .
\label{eq 21}
\end{equation}
Substituting into Eq.(\ref{eq 19}) leads to
\begin{equation}
\zeta_{0\tau_5} = \alpha_2 \left[ \zeta_{0(5\xi)} + \frac{5}{3}
\left(\frac{\beta_1}{\alpha_1}\right) \zeta_0 \zeta_{0\xi\xi\xi} +
\frac{10}{3} \left(\frac{\beta_1}{\alpha_1}\right)
\zeta_{0\xi} \zeta_{0\xi\xi} + \frac{5}{6} \left(\frac{\beta_1}{\alpha_1}
\right)^2 \zeta_0^2 \zeta_{0\xi} \right] \, ,
\label{eq 22}
\end{equation}
which is the $5^{th}$ order equation of the KdV hierarchy.
Different choices of the free parameter $\alpha_2$ correspond to
different normalizations of the time $\tau_5$. In particular, for
$\alpha_2 = 6$ it acquires the canonical form \cite{dodd}
\begin{equation}
\zeta_{0\tau_5} = 6 \zeta_{0(5\xi)} + 10 \alpha \zeta_0 \zeta_{0\xi\xi\xi} +
20 \alpha \zeta_{0\xi} \zeta_{0\xi\xi} + 5 {\alpha}^2 \zeta_0^2 \zeta_{0\xi}
\, ,
\label{eq 23}
\end{equation}
with $\alpha = \beta_1/\alpha_1$.

This procedure can be extended to any higher order. In other words, if
$\zeta_0$ satisfies the KdV equation in the time $\tau_3$, by making use of
the scale invariance described above, we can first select all possible terms
appearing in the evolution of $\zeta_0$ in the time $\tau_{2n+1}$. Then, by
imposing the general time symmetry condition
\begin{equation}
\frac{\partial^2 \zeta_0}{\partial \tau_{3} \partial \tau_{2n+1}} = \frac{
\partial^2 \zeta_0}{\partial \tau_{2n+1} \partial \tau_3} \, ,
\label{eq 26}
\end{equation}
the parameters $\alpha_n, \beta_n, \gamma_n, \ldots$ \, appearing in that
equation can be uniquely determined in terms of $\alpha_1$ and $\beta_1$, and
the resulting equation is found to be the $(2n+1)^{th}$ order equation of the
KdV hierarchy.
As before, $\alpha_n$ is left as a free parameter responsible for the
arbitrariness in the scale of the time $\tau_{2n+1}$. One gets easily convinced
of the validity of this
result by performing  the calculation  explicitly for the first few equations,
but we do not intend here
to give a general proof valid for any order.

Before proceeding further, let us remark that one could {\it a priori}
envisage, in place of Eq.(62) a new one
including also a dependence on $\zeta_1$, and still keep the agreement with
the scale invariance requirement.
However, when introducing this back into the physical equation (60), the
compatibility  condition (63)
makes all the supplementary terms to vanish.

\section{Higher Order Evolution Equations: the Elimination of the Secular
Producing Terms}

Let us now return to the evolution equation for $\zeta_1$ in the time $\tau_3$,
which is given by Eqs.(\ref{eq 72}) and (\ref{eq 73}). As we have seen, there
were two problems related to it: the evolution of $\zeta_0$ in $\tau_5$, and
the secular producing term. The last section gave the clue to the first
problem. As we have seen, the evolution of $\zeta_0$ in $\tau_5$ can not be
chosen arbitrarily, but it is restricted to the KdV hierarchy equations. We
show now how the secular producing term $\zeta_{0(5\xi)}$ can be eliminated.
A comparison between Eq.(\ref{eq 73}) for $S(\zeta_0)$ and the $5^{th}$ order
KdV equation (\ref{eq 22}), immediately shows that the choice
\begin{equation}
\alpha_2 = - \frac{19}{360} c h^4
\label{eq 74}
\end{equation}
has the property of eliminating the terms
\begin{equation}
\pm \zeta_{0\tau_5} - \frac{19}{360} c h^4 \zeta_{0(5\xi)}
\label{eq 75}
\end{equation}
from the non--homogeneous term $S(\zeta_0)$. Consequently,
$S(\zeta_0)$ acquires the form
\begin{equation}
S(\zeta_0) = \frac{9}{24} c h \zeta_0 \zeta_{0\xi\xi\xi} +
\frac{5}{8} c h \zeta_{0\xi} \zeta_{0\xi\xi} + \frac{63}{16}
\frac{c}{h^2} {\zeta_0}^2 \zeta_{0\xi} \, ,
\label{eq 76}
\end{equation}
and we see that it does not present a secular producing term anymore.

It should be noted that the resonant term is the linear term of $S(\zeta_0)$.
This property holds in any higher order of the perturbative scheme \cite{kota}.
Therefore, the elimination of the secular producing term by choosing an
appropriate value for t
he free parameter $\alpha_n$ in the higher--order equation of the KdV
hierarchy, also remains possible in any higher order.

\section{Traveling Wave Solutions: the Renormalization of the Soliton Velocity}

We have already assumed, in the context of the multiple time formalism, that
\begin{equation}
\zeta_0 = \zeta_0(\xi;\tau_3,\tau_5,\tau_7,\ldots)\, ,
\label{eq 77}
\end{equation}
where
\begin{equation}
\xi = {\xi}^\prime - c \tau_1 \, ,
\label{eq 78}
\end{equation}
so that $\zeta_0$ automatically satisfies the linear wave equation
\begin{equation}
\zeta_{0\tau_1} + c \zeta_{0\xi} = 0 \, .
\label{eq 79}
\end{equation}
Our concern now will be the solutions to the higher order equations of the KdV
hierarchy. Hereafter, for simplicity, we will assume that $c$ and $h$ can be
set equal to unity, so that the KdV equation reads
\begin{equation}
\zeta_{0\tau_3} = \alpha_1 \zeta_{0\xi\xi\xi} + \beta_1 \zeta_0 \zeta_{0\xi} \,
,
\label{eq 80}
\end{equation}
with
\begin{equation}
\alpha_1 = \frac{1}{6} \quad , \quad \beta_1 = \frac{3}{2} \, .
\label{eq 81}
\end{equation}
We now look for a traveling wave solution $\zeta_0$ that satisfies this
equation in the time $\tau_3$, but that satifies also the higher order
equations of the KdV hierarchy in the times $\tau_5,
\tau_7, \ldots \;\;$. This {\it multi--solution} can be written in the form
\begin{equation}
\zeta_0 = \frac{k^2}{3} {\rm sech}^2 \left(k \Lambda\right)
\label{eq 82}
\end{equation}
with the argument $\Lambda$ given by
\begin{equation}
\Lambda = \xi - \frac{k^2}{A_1} \tau_3 - \frac{k^4}{A_2} \tau_5 -
\frac{k^6}{A_3} \tau_7 - \ldots \, ,
\label{eq 83}
\end{equation}
and with $A_1, A_2, A_3, \dots \;$ parameters depending respectively on the
constants $\alpha_1$, $\alpha_2$,
$\alpha_3,~\ldots \;$. This argument is to be interpreted in the following way:
when the evolution equation under consideration is concerned to the time
$\tau_{2n+1}$, the relevant argument is
$$
\Lambda = \xi - \sum_{i=1}^{n} \frac{k^{2i}}{A_i} \tau_{2i+1}
- \theta_n \, ,
$$
with $\theta_n$ a phase involving all the remaining terms of the sum.
Then, for the KdV equation~(\ref{eq 80}), $A_1 = (1/\alpha_1) = 6$,
and the solution is
\begin{equation}
\zeta_0 = \frac{k^2}{3} {\rm sech}^2 \left[k \left(\xi - \frac{k^2}{6}
\tau_3 - \theta_3(k;\tau_5,\ldots) \right) \right] \, ,
\label{eq 84}
\end{equation}
where $\theta_3$ is a phase. On the other hand, the travelling wave solution
satisfying simultaneously the KdV equation~(\ref{eq 80}) in the time $\tau_3$,
and the $5^{th}$ order KdV equation~(\ref{eq 22}) in the time $\tau_5$, is
given by
\begin{equation}
\zeta_0 = \frac{k^2}{3} {\rm sech}^2 \left[k \left(\xi - \frac{k^2}{6}
\tau_3 - \frac{k^4}{A_2} \tau_5 - \theta_5(k;\tau_7,\ldots) \right)
\right] \, ,
\label{eq 85}
\end{equation}
and so on, to any higher order.

Once we have found the solutions to any member of the KdV hierarchy, let us
return to the equation for $\zeta_1$ in the time $\tau_3$, which is now $(c = h
= 1)$ written in the form
\begin{equation}
\zeta_{1\tau_3} + \frac{3}{2} {\left(\zeta_0
\zeta_1\right)}_{\xi} + \frac{1}{6} \zeta_{1\xi\xi\xi} = S(\zeta_0)
\, ,
\label{eq 86}
\end{equation}
with
\begin{equation}
S(\zeta_0) = \frac{9}{24} \zeta_0 \zeta_{0\xi\xi\xi} + \frac{5}{8}
\zeta_{0\xi} \zeta_{0\xi\xi} + \frac{63}{16} {\zeta_0}^2 \zeta_{0\xi}
\, .
\label{eq 87}
\end{equation}
It is important to remember that the linear term of $S(\zeta_0)$, which is
the secular producing term in the equation for $\zeta_1$, was eliminated
through the use of the $5^{th}$ order KdV equation~(\ref{eq 22}), with
$\alpha_2$ given by Eq.(\ref{eq 74}).
Therefore, the $\zeta_0$ appearing in Eqs.(\ref{eq 86}--\ref{eq 87}) is that
given by Eq.(\ref{eq 85}), since it must be a solution of KdV as well as of the
KdV hierarchy $5^{th}$ order equation
(\ref{eq 22}).
The secular--free solution $\zeta_1$ can then be found by solving the linear
equation (\ref{eq 86}--\ref{eq 87}). This solution can be found in
Ref.\cite{kota}.

Finally, let us take the solutions $\zeta_0$, and let us make the
transformation back from the slow $(\xi,\tau,\zeta_0)$ to the laboratory
coordinates $(x,t,\zeta)$. For the case of solution (\ref{eq 84}) to the KdV
equation, we get
\begin{equation}
\zeta = \frac{k^2}{3} \epsilon \, {\rm sech}^2 \left[ k \, \epsilon^{1/2}
\left( x - V_3 \, t \right)\right] \, ,
\label{eq 88}
\end{equation}
where
\begin{equation}
V_3 = c + \epsilon \, \frac{k^2}{6}
\label{eq 89}
\end{equation}
is the solitary--wave velocity in the laboratory coordinates. For the case of
solution (\ref{eq 85}) to the $5^{th}$ order KdV equation, we get
\begin{equation}
\zeta = \frac{k^2}{3} \epsilon \, {\rm sech}^2 \left[ k \, \epsilon^{1/2}
\left( x - V_5 \, t \right)\right] \, ,
\label{eq 90}
\end{equation}
where now
\begin{equation}
V_5 = V_3 + \epsilon^2 \, \frac{k^4}{A_2}
\, .\label{eq 91}
\end{equation}
Since $A_2$ depends on the parameter $\alpha_2$, to choose $\alpha_2$ means to
define the velocity renormalization itself. In this sense we can say that there
is a unique velocity renormalization leading to a secular--free perturbation
scheme for $\zeta_1
$. From these properties, we can see now that this method is equivalent to the
renormalization technique developed by Kodama and Taniuti \cite{kota}, in which
the secular--free higher order effects were also given by the renormalization
of the KdV soliton
velocity. Moreover, in the same way as in the method of Kodama and Taniuti, if
higher order scales are introduced, it is possible to continue the
secular--free perturbation to higher orders by using the higher order equations
of the KdV hierarchy. And in
 general, depending on the order considered in the perturbative scheme, the
renormalized solitary--wave velocity in the laboratory coordinates is given by
\begin{equation}
V_{2n+1} = c + \sum_{i=1}^{n} \epsilon^i \, \frac{k^{2i}}{A_i} \, .
\label{eq 92}
\end{equation}
The analysis for multi--soliton solutions proceeds in a similar fashion as
well.

\section{Final Remarks}

By using the reductive perturbation method of Taniuti \cite{taniu}, with the
introduction of an infinite sequence
of slow time variables $\tau_1$, $\tau_3$, $\tau_5,~\ldots \; $,  we studied
the propagation of long surface waves in a shallow inviscid fluid. The three
main ingredients of our
analysis were: (i) the scale--invariance argument, which restricts the possible
forms of the evolution equations, and which
is necessary for the coherence of the perturbative expansion; (ii) the
compatibility condition (63), which
appears when  extending  a function of a single time-variable to a multiple
time--variable; (iii) the secularity elimination procedure, without which the
perturbative expansion would be meaningless. By using (i)
and (ii) we have shown that, if the amplitude $\zeta_0$ satisfies the KdV
equation, it also satisfies the higher--order
equations of the KdV hierarchy. Then, by using the $5^{th}$ order  equation  of
the hierarchy with a
properly chosen $\alpha_2$, we have shown that the secular producing terms of
the equation for $\zeta_1$  in
the time $\tau_3$ could be eliminated, a result which can be extended up to any
higher order. As the secularity elimination is
mandatory for a physical theory, being essencially a finiteness requirement,
the results obtained in this paper allowed us to give a
physical meaning to the KdV hierarchy equations. Thereafter, by considering a
solitary wave solution, we have shown that the elimination of
secularities through the use of
higher order KdV equations corresponds, in the laboratory coordinates, to a
renormalization of the
soliton velocities, as obtained previously by Kodama and Taniuti \cite{kota}.

The study of long--waves in shallow water is representative of a wide class,
that of the weak
nonlinear, dispersive systems, where the KdV equation has a kind of universal
character. In this sense, we can say that
the results of the present paper do not depend on the specific physical system
under consideration, or, which is the same, on the specific
form of the basic equations, except for the values of the coefficients in the
perturbative expansion. It
is, therefore, legitimate to conjecture that they might be extended to the
above mentioned larger class of systems.

\vspace{1 cm}
\section*{Acknowledgements}

The authors would like to thank J. L\'eon for usefull
discussions. Two of the authors (RAK and JGP), would like to
thank CNPq, Brazil, for partial support. RAK also thanks CAPES, Brazil, for
financial support.

\end{document}